\documentstyle[aps,eqsecnum,preprint,prd,epsf]{revtex}

\tighten
\begin{document}
\draft

\preprint{TUM--T31--54/93}
\title{Finite Mass Corrections to Leptonic Decay Constants in the\\ Heavy
Quark Effective Theory}
\author{Patricia Ball}
\address{Physik--Department/T30, TU M\"unchen, D--85747 Garching, FRG}
\date{November 28, 1993}
\maketitle
\begin{abstract}
We calculate finite mass corrections to leptonic decay constants in HQET
using QCD sum rules. The results are given to two--loop accuracy
including renormalization group analysis. The contribution of unknown two--loop
anomalous dimension matrix--elements is estimated.
We find that the $1/m_Q$ corrections to $f_P$, the decay constant of
a pseudoscalar meson, amount 
$-(23\pm4\pm2)\%$ for the B meson. The first error comes from the
sum rule uncertainty, the second one from unknown or not accurately
known matrix elements in HQET. We obtain a ratio of vector to pseudoscalar
decay constant of $1.00\pm0.07+O(1/m_Q^2)$ for B mesons.
\end{abstract}
\clearpage

\section{Introduction}

QCD in the limit of infinite heavy quarks has received much interest
over the past years (cf.\ \cite{reviews} and references therein). 
The theory simplifies considerably in that limit and though it still
does not allow a complete calculation of hadronic matrix elements,
it poses some conditions on and gives relations between them that can
be used to reduce the model--dependence of the analysis of experimental data. 
Its phenomenological relevance is
however limited by the fact that quark masses are finite. Generally,
matrix elements determined in the heavy quark effective theory (HQET) 
receive corrections due to the finite quark mass. Since we are
dealing with a field theory that has to be renormalized, the
corrections are not pure powers, but are modified by
large logarithms of type $\ln m_Q/\mu$ where $m_Q$ is the heavy quark
mass and $\mu$ some hadronic scale. Physically, the logarithms are
due to the radiation of high energetic gluons off the heavy quark.  
Although in certain cases some terms in the expansion can be
shown to be small or absent (as is the case with Luke's theorem 
\cite{Luke}), in general they will introduce new unknown quantities
order by order in the heavy quark expansion. If HQET is supposed to
yield results of phenomenological relevance, it is thus of paramount
importance to keep those corrections under control.

In the present paper we address the question of $1/m_Q$ corrections to
one specific matrix element, the leptonic decay constant $f_M$ of a
pseudoscalar meson $M$ with momentum $p$, built of the heavy quark $Q$ 
and the light anti--quark $\bar q$:
\begin{equation}\label{eq:deffM}
\langle \, 0 \, | \, \bar q \gamma_\mu \gamma_5 Q \, | \,M(p) \, \rangle
= if_Mp_\mu.
\end{equation}
From the analysis of potential models \cite{potmod}, it was known long since
that $F\equiv f_M \sqrt{m_Q}$ approaches a constant as $m_Q$ becomes
infinitely heavy. The next
step in unravelling the mysteries of $f_M$ was the
discovery that the simple square--root like behaviour gets modified by
logarithmic terms \cite{volshif}. After the
development of the formal framework of heavy quark QCD as
an effective theory \cite{georgieff}, $F$ was soon subject to
numerical studies on the lattice \cite{fBlatt} and by QCD sum rules 
\cite{BBBD,NA,N92}.
After some fluctuations, the results now agree within the errors:
\begin{equation}
F(\mu=m_B) \approx (0.4-0.6) \,\text{GeV}^{3/2}.
\end{equation}
The value of $f_B$, the decay constant of the B meson, that can be
derived from the above value is however considerably bigger than values
obtained in ``full QCD'' without reference to HQET \cite{fBlatt,fBQCD}.
Since $f_B$ is an important quantity in many applications like
$B_0$--$\bar B_0$ mixing, e.g.\ (cf.\ \cite{ECFA}), its knowlegde is 
of great importance and thus
requires a careful analysis of finite mass corrections to $F$.
According to estimates of various authors, $1/m_Q$ corrections to the
heavy quark limit amount to about 20\% \cite{fBlatt,BBBD,fBSR1overm}. 

In a previous paper \cite{BBBD} we have treated higher order corrections 
based on a ``higher twist'' expansion of a QCD sum rule for
$f_B$. An explicit expansion in inverse powers of $m_Q$ did not
prove feasible since the expansion of all parameters inherent in the sum
rule analysis in
inverse powers of $m_Q$ is by no means unique. Instead we defined the
``correction term'' to be the difference between QCD and HQET result.
In the range $m_Q> 3\,\text{GeV}$ the corrections proved to be
dominated by the linear term in the heavy quark expansion. On the
other hand, in \cite{N92} an attempt was made to compute $1/m_Q$
directly in HQET, but still relied heavily on the expansion of sum
rule parameters in $1/m_Q$. The resulting $1/m_Q$ terms are about
a factor of three larger than the correction obtained in
\cite{BBBD}. Thus we feel the
necessity to complement our old calculation by a corresponding one where
the $1/m_Q$ terms are investigated entirely in HQET. Moreover, such an
calculation offers the possibility to include next--to--leading order
renormalization group effects studied recently in \cite{N93}.

Our paper is organized as follows: in \mbox{Sec.\/ II} we recall the
formal framework of HQET and give necessary definitions, \mbox{Sec.\/
III} deals with the heavy quark expansion of correlation functions, in
\mbox{Sec.\/ IV} we attempt an approximate non--perturbative
calculation of the relevant matrix elements in HQET by QCD sum rules.
The last section, \mbox{Sec.\/ V}, is devoted to the presentation of
the results and their discussion.

\section{Preliminaries and Definitions}

Before starting with ``heavy'' calculations, let us recall the
main features of HQET. The heavy quark field $h_v$ is defined by
\begin{equation}
P_+ Q(x) = e^{-im_Q(v\cdot x)} h_v(x),
\end{equation}
where 
\begin{equation}
P_+ = \frac{1}{2} (1+\rlap/v)
\end{equation}
is the projector on the upper components of the heavy quark field
$Q(x)$, $m_Q$ is the mass of the heavy quark, and $v$ its four--velocity.
The contribution of the lower components to
the QCD Lagrangian can be integrated out
and the new effective Lagrangian be expanded in inverse powers of
the heavy quark mass\footnote{By the heavy quark mass $m_Q$ we
understand the pole mass whose precise definition will be given in 
(\ref{eq:polemass}).} \cite{FGL91,MRR92}:
\begin{equation}\label{eq:Lagrangian}
{\cal L}_{HQET} = \bar{h}_v i(v\!\cdot\! D)h_v + \frac{1}{2m_Q}\,{\cal K} +
\frac{1}{2m_Q}\, C_{\text{mag}}(\mu){\cal S}+ O(1/m_Q^2).
\end{equation}
At order $1/m_Q$, there are two operators,\footnote{A third one,
$\bar{h}_v (iv\!\cdot\! D)^2 h_v$, can be shown to be of higher order
in $1/m_Q$ by virtue of the equation of motion of the heavy quark
field.} the kinetic energy operator
\begin{equation}
{\cal K} = \bar{h}_v (iD)^2 h_v
\end{equation}
whose matrix element over hadron states with
equal velocities is related to the kinetic energy of the heavy quark
within these states. Due to reparametrization invariance
${\cal K}$ does not get renormalized \cite{LM92}, whereas the
chromomagnetic interaction operator,
\begin{equation}
{\cal S} = \frac{1}{2} \bar{h}_v\sigma_{\mu\nu}gF^{\mu\nu}h_v
\end{equation}
acquires an anomalous dimension $\gamma^{\text{mag}} = 6 \alpha_s/(4\pi) +
O(\alpha_s^2)$ \cite{FGL91} that gives rise to the Wilson coefficient
$C_{\text{mag}}(\mu)$.

In HQET, the heavy quark does not take part in the dynamics of a
process that is essentially determined by the hadron's light degrees 
of freedom. As a consequence, the heavy
quark spin becomes a ``good'' quantum number and the theory exhibits
the famous spin--symmetry\footnote{Of course the
symmetry breaks down at order $1/m_Q$, but in a so to speak controlled
manner that is induced by ${\cal S}$ and only affects the traces over
spin wave functions.} which leads to a factorization of
matrix elements into terms describing the spin--structure of the
involved hadrons (and their interaction) and terms describing the
dynamics of the process that is independent of the heavy quarks. The formal
tool to handle that factorization is the so--called trace--formalism 
\cite{traceformalism}. In this formalism and using the notation of
\cite{N92}, we define, making explicit the dependence on the
renormalization scale $\mu$ whenever necessary,
\begin{equation}\label{eq:defF}
\langle \, 0 \, | \, \bar{q}\Gamma h_v \, | \, M\,\rangle =
\frac{1}{2}\,F(\mu)\,\text{Tr}\{ \Gamma {\cal M}\}.
\end{equation}
${\cal M}$ is the spin wave function of the HQET meson state
$|\,M\,\rangle$:
\begin{equation}
{\cal M} = \sqrt{m_Q}\,P_+ \left\{ \begin{array}{cl}
-i \gamma_5 & \text{for\ } J^P = 0^-,\\
\not\! \epsilon & \text{for\ } J^P = 1^-.
\end{array}\right.
\end{equation}
Here $\epsilon_\mu$ is the polarization vector of the vector meson.

At order $1/m_Q$ and neglecting radiative corrections, the leptonic 
decay constant $f_M$ is determined by (cf.\ \cite{N92})
\begin{equation}\label{eq:deltaF}
f_M\,\sqrt{m_M} = F\left(1-d_M\frac{\bar\Lambda}{6m_Q} + 
\frac{G_K}{2m_Q}+d_M\,\frac{G_\Sigma}{m_Q}\right).
\end{equation}
$\bar\Lambda$ is defined as the mass--difference of meson mass $m_M$
and quark mass $m_Q$ in the heavy quark limit, $m_M = m_Q + 
\bar\Lambda+O(1/m_Q)$. $G_K$ and $G_\Sigma$ are defined by
\begin{eqnarray}
\langle \,0\,|\,i\!\!\int\!\! d^4x\, {\cal K}(x)\,\bar{q}(0)\Gamma
h_v(0)\, |\, M\,\rangle & = & \frac{1}{2}\,F(\mu) G_K(\mu)
\text{Tr}\{\Gamma{\cal M}\},\nonumber\\
\langle \,0\,|\,i\!\!\int\!\! d^4x\, {\cal S}(x)\,\bar{q}(0)\Gamma
h_v(0)\, |\, M\,\rangle & = & \frac{1}{2}\,F(\mu) G_\Sigma(\mu) 2d_M
\text{Tr}\{\Gamma{\cal M}\}.\label{eq:defG}
\end{eqnarray}
As already mentioned, at order $1/m_Q$ the spin--symmetry breaks down
and matrix elements become dependent on the spin of the heavy
particle as characterized by a scalar $d_M$ defined by
\begin{equation}
P_+\sigma_{\alpha\beta}{\cal M}\sigma^{\alpha\beta} = 2 d_M {\cal
M},
\end{equation}
yielding
\begin{equation}
d_M = \left\{\begin{array}{cl}
3 & \text{for\ }J^P = 0^-,\\
-1 & \text{for\ }J^P = 1^-.
\end{array}
\right.
\end{equation}

Eq.\/ (\ref{eq:deltaF}) contains all information about $f_M$ to order
$1/m_Q$ and to lowest order in the strong coupling, but involves the 
three new matrix
elements $\bar\Lambda$, $G_K$ and $G_\Sigma$. For $\bar\Lambda$ there 
exists a rigorous lower bound of
$237\,\text{MeV}$ \cite{guralnik} that is however questioned in
\cite{bigi}. Neglecting $1/m_Q$ effects, the
values of the b--quark mass quoted in literature,
$m_b = (4.6-4.8)\,\text{GeV}$ \cite{mb}, yield $\bar\Lambda =
(0.5-0.7)\,\text{GeV}$. More refined analyses in HQET \cite{BBBD,N92}
tend to the somewhat smaller value $\bar\Lambda =
(0.5-0.6)\,\text{GeV}$ which we will use in our paper.

The available values of $G_K$ and $G_\Sigma$ rely exclusively on the
analysis given in \cite{N92}
that is similar in spirit to the one we are going to undertake but
quite different in detail. We will come back to that point later. The
results of \cite{N92} read (at a renormalization scale of 
$5\,\text{GeV}$):
\begin{equation}
G_K \simeq -4.4 \,\text{GeV},\quad G_\Sigma \simeq -0.05\,\text{GeV}.
\end{equation}

\section{The heavy quark expansion of correlation functions}

Our calculation of $G_K$ and $G_\Sigma$ relies on the analysis of
two--point functions of the heavy--light current $J_\Gamma = \bar{q}
\Gamma h_v$ with insertions of ${\cal K}$ and ${\cal S}$,
respectively. Since we will include radiative corrections to
$O(\alpha_s)$, an investigation of the renormalization--group
behaviour proves necessary. In addition,
at the required level of accuracy we need to match the effective
theory to QCD, i.e.\/ we have to require the equality of matrix
elements in HQET to the corresponding ones in QCD at the
matching scale $\mu = m_Q$. As a by--product of this procedure, we will
be able to check the heavy quark expansion explicitly at the two--loop
level at the example of the two--point function of pseudoscalar
currents. This will be done in the next section.

Let us first introduce a generic notation for correlation functions of
two composite operators ${\cal O}_1$ and ${\cal O}_2$, either containing
a heavy quark field,
\begin{equation}\label{eq:notation}
\langle\, {\cal O}_1 {\cal O}_2\,\rangle \equiv i\!\!\int\!\!
d^4x\,e^{i\omega v\cdot x}
\, \langle\, 0\,|\, {\bf T} {\cal O}_1(x) {\cal O}_2^\dagger(0)\, |\,
0\,\rangle,
\end{equation}
with four--momentum $\omega v$, $v$ being the velocity of the
heavy quark. Analogously, we define for the
zero--momentum insertion of the operator ${\cal X}$:
\begin{equation}\label{eq:insertion}
\langle\, {\cal O}_1 {\cal X} {\cal O}_2\, \rangle \equiv i^2\!\!\int\!\!
d^4x\, d^4y\, e^{i\omega v\cdot (x-y)}\, \langle\,0\,|\,{\bf T}{\cal
O}_1(x) {\cal X}(0) {\cal O}_2^\dagger(y)\,|\,0\,\rangle.
\end{equation}
Employing the trace--formalism for heavy--light currents $J_{\Gamma_i}
= \bar q \Gamma_i h_v$ we can write\footnote{Our definition deviates 
formally from the conventional one $\langle\,
J_{\Gamma_1} J_{\Gamma_2}\,\rangle = -\frac{1}{2}\,\Pi(\omega)\,
\text{Tr}(\Gamma_1P_+\Gamma_2)$, but coincides for ``spin
eigenstates'' like $\Gamma_i = i\gamma_5$ or $\Gamma_i =\gamma_\mu-
v_\mu$ with \mbox{$\Gamma_i \rlap/ v = - \rlap/ v \Gamma_i$} and can
be directly inferred from the structure of the relevant loop--integrals.}
\begin{equation}
\langle J_{\Gamma_1} J_{\Gamma_2}\,\rangle =
\frac{1}{2}\,\Pi(\omega,\mu)\,\text{Tr}\{\Gamma_1P_+\Gamma_2\rlap/ v\}.
\end{equation}
Recall that $\Pi(\omega,\mu)$ is independent of both $\Gamma_1$ and
$\Gamma_2$, but depends on the renormalization scale $\mu$ due to the
non--vanishing anomalous dimension of $J_\Gamma$. With ${\cal X} =
{\cal K}$ or ${\cal S}$, (\ref{eq:insertion}) can be expressed as
\begin{eqnarray}
\langle\, J_{\Gamma_1}{\cal K}J_{\Gamma_2}\, \rangle & \equiv &
T_K(\omega,\mu)\,\text{Tr}\{\Gamma_1 P_+\Gamma_2\rlap/ v\},\nonumber\\
\langle\, J_{\Gamma_1}{\cal S}J_{\Gamma_2}\, \rangle & \equiv &
T_\Sigma(\omega,\mu)\,d_M\,\text{Tr}\{\Gamma_1 P_+\Gamma_2\rlap/ v\}.
\end{eqnarray}

In QCD, the renormalization--group invariant two--point function of 
pseudoscalar currents with momentum $q$ is defined as
\begin{equation}\label{eq:Pi5}
\Pi_5 = i\!\!\int\!\! d^4x\,e^{iqx}\, \langle\,0\,|\,{\bf T}\,
m_{\overline{\text{MS}}}\, \bar{q}(x)i\gamma_5 Q(x)\,
m_{\overline{\text{MS}}}\, \bar{Q}(0)i\gamma_5q(0)\,|\,0\,\rangle.
\end{equation}
Here the running $\overline{\text{MS}}$ mass
$m_{\overline{\text{MS}}}$ 
enters whose relation to the pole mass is given by
\begin{equation}\label{eq:polemass}
m_Q = m_{\overline{\text{MS}}}(\mu)\left[1+\frac{\alpha_s(\mu)}{\pi}
\left( \frac{4}{3} - 2 \ln \frac{m_Q}{\mu}\right) + O\left( \alpha_s^2
\right)\right].
\end{equation}
$\Pi_5$ was calculated to two--loop accurcay in the deep Euclidean 
region $q^2\ll 0$ in \cite{broadhurst} for arbitrary quark masses. Once
the heavy quark expansion of the pseudoscalar current is known, the
expression for $\Pi_5$ can serve as check for both the cancellation of
$\mu$ dependent terms
in the effective theory and the values of the matching coefficients,
i.e.\/ the values of the short--distance expansion coefficients at
$\mu = m_Q$.

As for the pseudoscalar current, its expansion reads (following the
notation of \cite{N93}):
\begin{equation}\label{eq:expPS}
m_{\overline{\text{MS}}} \bar q i\gamma_5 Q \cong m_Q\left(C(\mu) 
J_{i\gamma_5} + \frac{1}{2m_Q}\, \sum_{i}^2 B_i(\mu) {\cal O}_i + 
\frac{1}{2m_Q}\,\sum_k^{2} A_k(\mu) T_k\right).
\end{equation}
The above formula needs some explanation. First, ``$\cong$'' means
equality only after taking matrix elements or insertion in Greens
functions. Next, we have kept the multiplicative mass term in order to
indicate that it affects the
coefficient functions on the right hand side by virtue of Eq.\/ 
(\ref{eq:polemass}). The leading order operator in HQET is the current
$J_{i\gamma_5}$; at order $1/m_Q$, there are two local operators,
\begin{equation}
{\cal O}_1 = \bar{q}i\gamma_5i\,\rlap/\! D h_v, \quad {\cal O}_2 =
\bar{q}(-iv\! \cdot\! \overleftarrow{D})i\gamma_5 h_v,
\end{equation}
and two non--local ones due to the insertion of the Lagrangian into
the leading term:
\begin{equation}
T_1 = i\!\!\int\!\! d^4x\, {\bf T}J_{i\gamma_5}(x) {\cal K}(0), \quad
T_2 = i\!\!\int\!\! d^4x\, {\bf T}J_{i\gamma_5}(x) {\cal S}(0).
\end{equation}
The operators have non--vanishing anomalous dimensions and mix
under renormalization. The mixing is described by an $4\times 4$
matrix $\hat{\gamma}^{\text{PS}}$ and 
the coefficient functions $\{B_1,B_2,A_1,A_2\} \equiv \vec{C}$ obey the 
renormalization--group equation
\begin{equation}\label{eq:renWC}
\left( \mu\,\frac{d}{d\mu} - \hat{\gamma}^{\text{PS}}\right) \vec{C} =
0.
\end{equation} 
Since ${\cal K}$ does
not get renormalized, it follows $A_1(\mu) = C(\mu)$. For
$A_2$, we have $A_2(\mu) = C(\mu) C_{\text{mag}}(\mu)$ due to the
non--zero anomalous dimension of ${\cal S}$ (cf.\/
(\ref{eq:Lagrangian})). $C_{\text{mag}}$ was calculated in
\cite{FGL91,EH90} to be
\begin{equation}
C_{\text{mag}}(\mu)= 1 + \frac{\alpha_s}{\pi}\left( \frac{13}{6} + 
\frac{3}{2}\,\ln \frac{m_Q}{\mu} \right).
\end{equation}

In order to determine the remaining coefficients, we proceed like
Ref.\/ \cite{N93} and calculate matrix--elements between on--shell
quark states. The corresponding diagram is shown in 
\mbox{Fig.\/ \ref{fig:momconf}}. The on--shell spinors
obey the Dirac equations $\rlap/  p u_Q(p) = m_Q u_Q(p)$ 
and $\rlap/  p' u_q(p') = 0$. The heavy quark momentum $p$ can be
written as $p = m_Q v + k$ where the off--shell momentum $k$ obeys 
$2 m_Q v\!\cdot \! k + k^2=0$ in order to keep $u_Q$ on--shell. For
the on--shell spinor in HQET the Dirac equation reads
$\rlap/ v u_h=0$, and it follows the relation $u_Q = \{1+ \rlap/
k/(2m_Q)\} u_h$. Taking into account only terms constant or linear in
$p'$, loop--integrals in HQET vanish in dimensional
regularization since there is no massive scale they could depend on.
The diagram in Fig.\/ \ref{fig:momconf}, wave function renormalization
and terms coming from the replacement of the running
$\overline{\text{MS}}$ mass by the pole mass yield\footnote{We use 
anticommuting $\gamma_5$.}
\begin{equation}\label{eq:PScoeff}
C(\mu) = B_1(\mu) = 1+\frac{\alpha_s}{\pi}\left( \ln\,
\frac{m_Q}{\mu}-\frac{2}{3}\right), \quad B_2(\mu) =
\frac{8}{3}\,\frac{\alpha_s}{\pi}.
\end{equation}
The expression for $C(\mu)$ is well known and can be found in a
somewhat different form in \cite{JM} apart from a slight error that
was corrected in \cite{BBBD,NA}. The first equality in
(\ref{eq:PScoeff}) is consistent with the requirement of
reparametrization invariance, the expression for $B_2$ is new.

The expansion of (\ref{eq:Pi5}) to order $1/m_Q$ is now
straightforward to derive. Using
\begin{equation}
\langle\, J_{\Gamma_1 i D_\alpha} J_{\Gamma_2} \,\rangle =
\frac{1}{6}\, \omega \Pi(\omega)\,\text{Tr}\{(\gamma_\alpha-v_\alpha
\rlap/ v) \Gamma_1P_+\Gamma_2\},
\end{equation}
the insertion of (\ref{eq:expPS}) into (\ref{eq:Pi5}) yields
\begin{eqnarray}
\Pi_5 & =  & m_Q^2\left(C^2(\mu) \Pi(\omega,\mu) + \frac{1}{m_Q}\left[
\vphantom{\frac{1}{m_Q}}\left\{C^2(\mu) - 
B_2(\mu) C(\mu)\right\} \omega \Pi(\omega,\mu) + C^2(\mu)
T_K(\omega,\mu)\right.\right.\nonumber\\
& & {}\left.\left. + 3 C^2(\mu) C_{\text{mag}}(\mu) T_\Sigma(\omega,\mu)
\right]\right).\label{eq:expansion}
\end{eqnarray}
This expression is valid up to constants in $\omega$ that are related
to contact terms and cannot be expanded in $1/m_Q$. The 
$\mu$--dependent terms on the right hand side must cancel in order to
keep the result renormalization--group invariant. This cancellation
will be checked in the next section.

One could now proceed by solving the renormalization--group equation
(\ref{eq:renWC}) as was done in \cite{N92}. For our purposes, however,
it proves more appropriate to investigate the behaviour of the
two--point functions $\vec{\cal V} = \{\omega
\Pi(\omega), T_K(\omega), T_\Sigma(\omega)\}$ under
renormalization--group. $\vec{\cal V}$
obeys the renormalization group equation
\begin{equation}\label{eq:RG}
\left( \mu\, \frac{d}{d\mu} + \hat{\gamma}\right) \vec{\cal V}(\mu) =
0.
\end{equation}
The $3\times 3$ anomalous dimension matrix $\hat{\gamma}$ can be
obtained from the renormalization constant $\hat Z$ that relates the
bare $\vec{\cal V}_{\text{bare}}$ to the renormalized one, 
$\vec{\cal V}_{\text{bare}} = \hat{Z}\,\vec{\cal V}$. Restricting
ourselves to leading logarithmic accuracy (superscript LL), 
$\hat{\gamma}^{\text{LL}}$ can be obtained as
$\hat{\gamma}^{\text{LL}} = \lim_{\epsilon\to 0} 2\epsilon \hat{Z}$ in
the $\overline{\text{MS}}$ scheme\footnote{But note that the result is
renormalization--scheme independent to that accuracy.} using
dimensional regularization in $D=4+2\epsilon$ dimensions. For
$\omega\ll 0$, $\vec{\cal V}$ can be treated within perturbation theory
and we find
\begin{equation}\renewcommand{\arraystretch}{1.2}
\hat{\gamma}^{\text{LL}} = \frac{\alpha_s}{4\pi}\, \left(
\begin{array}{rrr}
-8 & 0 & 0\\
\frac{32}{3} & -8 & 0\\
-\frac{32}{9} & 0 & -2\\
\end{array}
\right) \equiv \frac{\alpha_s}{4\pi}\,\gamma.
\end{equation}\renewcommand{\arraystretch}{1}

To that accuracy, the solution of Eq.\ (\ref{eq:RG}) is
\begin{equation}
\vec{\cal V}(m_Q) = \exp\left( -\frac{\gamma}{2\beta_0}\,\ln x\right) 
\vec{\cal V}(\mu)
\end{equation}
with $x = \alpha_s(\mu)/\alpha_s(m_Q)$. $\beta = -g_s\left[\beta_0
\alpha_s/(4\pi) + \beta_1 (\alpha_s/(4\pi))^2+O(\alpha_s^2)\right]$ 
is the $\beta$ function of QCD. The matrix $\gamma$ cannot be
diagonalized, but transformed to its Jordan form $\gamma_J$ via
\begin{equation}
\gamma_J = T^{-1}\gamma T
\end{equation}
with
\begin{equation}
T  =  \left(
\begin{array}{rrr}
0 & -6 & 0\\
-64 & 1 & 0\\
0 & -\frac{32}{9} & 1\\
\end{array}
\right),\quad
\gamma_J = \left(
\begin{array}{rrr}
-8 & 1 & 0\\
0 & -8 & 0\\
0 & 0 & -2\\
\end{array}
\right).
\end{equation}
Using
\begin{equation}
\exp\left(
\begin{array}{cc}
a & 1\\
0 & a\\
\end{array}
\right)
=
\exp (a)\,\left(
\begin{array}{cc}
1 & 1\\
0 & 1\\
\end{array}
\right),
\end{equation}
we find
\begin{equation}\renewcommand{\arraystretch}{1.2}
\exp \left(-\frac{\gamma}{2\beta_0} \,\ln x\right) = x^{4/\beta_0} \left(
\begin{array}{ccc}
1 & 0 & 0\\
-\frac{16}{3\beta_0}\, \ln x & 1 & 0\\
-\frac{16}{27}\, (x^{-3/\beta_0} -1) & 0 & x^{-3/\beta_0}
\end{array}
\right).\renewcommand{\arraystretch}{1}
\end{equation}
As expected, ${\cal V}_1$ renormalizes like $\Pi(\omega)$. 
For ${\cal V}_2$ and ${\cal V}_3$ we can infer the renormalization--group
invariant quantities
\begin{eqnarray}
\hat{\cal V}_2^{\text{LL}} & \equiv & [\alpha_s(\mu)]^{4/\beta_0}\,
\left(-\frac{16}{3\beta_0}\, \ln \alpha_s(\mu)\, {\cal V}_1(\mu) +
{\cal V}_2(\mu)\right)\, ,\nonumber\\
\hat{\cal V}_3^{\text{LL}} & \equiv &
[\alpha_s(\mu)]^{1/\beta_0}\,\left(-\frac{16}{27} {\cal V}_1(\mu) +
{\cal V}_3(\mu)\right).
\end{eqnarray}
In order to extend the analysis to next--to--leading order
(superscript NLO), one needs
to include the anomalous dimension matrix $\hat{\gamma}$ to two--loop 
accurcay. For ${\cal V}_1$ one finds
\cite{BBBD,NA,BG92}
\begin{equation}
\hat{\cal V}_1^{\text{NLO}} =
[\alpha_s(\mu)]^{4/\beta_0}\,\left(1-
S_{\text{HL}}\,\frac{\alpha_s(\mu)}{2\pi}\right) {\cal V}_1(\mu),
\end{equation}
where 
\begin{equation}
S_{\text{HL}} = \frac{\gamma^{\text{HL}}_0}{2\beta_0} \left(
\frac{\gamma^{\text{HL}}_1}{\gamma^{\text{HL}}_0}
-\frac{\beta_1}{\beta_0}\right).
\end{equation}
$\gamma^{\text{HL}} = \gamma_0^{\text{HL}}\alpha_s/(4\pi) +
\gamma_1^{\text{HL}}\{\alpha_s/(4\pi)\}^2 + O(\alpha_s^3)$
is the anomalous dimension of the heavy--light current $J_\Gamma$
calculated in \cite{JM,BG91}. As for $\hat{\cal V}_2$ and $\hat{\cal
V}_3$, some of the involved elements of $\hat{\gamma}$ are still
unknown to date. According to \cite{N93} one obtains
\begin{eqnarray}
\hat{\cal V}_2 & = & [\alpha_s(\mu)]^{4/\beta_0}\,\left(1-
S_{\text{HL}}\,\frac{\alpha_s(\mu)}{2\pi}\right) \left\{V_2(\mu) - 
V_1(\mu) \left( \frac{16}{3\beta_0}\, \ln \alpha_s(\mu) +
\frac{8}{3}\, T_{34}\, \frac{\alpha_s(\mu)}{\pi} \right)\right\},\nonumber\\
\hat{\cal V}_3 & = & [\alpha_s(\mu)]^{1/\beta_0}\!\left(1-
2S_{\text{HL}}\,\frac{\alpha_s(\mu)}{2\pi}-S_{\text{mag}}\,
\frac{\alpha_s(\mu)}{4\pi}\right)\! \left\{V_3(\mu) - \frac{16}{27}\,
V_1(\mu)\! \left(1 + \frac{3}{2}\, T_{32}\,
\frac{\alpha_s(\mu)}{\pi} \right)\!\right\}.\label{eq:latest}
\end{eqnarray}
$S_{\text{mag}}$ is the analogue of $S_{\text{HL}}$ were $\gamma^{\text{HL}}$
has to be replaced by the anomalous dimension $\gamma^{\text{mag}}$ of
the operator ${\cal S}$. Unknown are the matrix elements $T_{32}$ and 
$T_{34}$ as well as $\gamma_1^{\text{mag}}$. 

We now want to relate the two--point functions we have been 
dealing with in this section to the $1/m_Q$ corrections to $F$ we are
interested in. To that end, we saturate the correlation function
$\Pi_5$ by hadronic states. Making explicit only the insertion of the
ground--state, the meson M, it follows with (\ref{eq:deffM})
\begin{equation}
\Pi_5 = \frac{m_M^4 f_M^2}{m_Q^2(m_M^2-q^2)} + \dots
\end{equation}
where the dots denote insertions of radial excitations and
multi--particle states.
Analogously, we find from (\ref{eq:defG}), choosing 
$\mu = m_Q$:
\begin{eqnarray}
\Pi(\omega,m_Q) & = & \frac{F^2(m_Q)}{2(\bar\Lambda-\omega)} +
\dots,\nonumber\\
T_K(\omega,m_Q) & = & \frac{F^2(m_Q) G_K(m_Q)}{2(\bar\Lambda-\omega)}
+ \dots ,\nonumber\\
T_\Sigma(\omega,m_Q) & = & \frac{F^2(m_Q) G_\Sigma(m_Q)
}{(\bar\Lambda-\omega)}+ \dots\label{eq:xyz}
\end{eqnarray}
For $T_K$ and $T_\Sigma$ the dots contain also additional double poles
at $\omega = \bar\Lambda$ that are related to $1/m_Q$ corrections to
$\bar\Lambda$ and are treated in detail in \cite{BB93}.

Identifying the leading terms in $1/m_Q$, we recover from
(\ref{eq:expansion}) the well known relation for the pseudoscalar
decay constant $f_P$ \cite{JM}
\begin{equation}
f_P^2 m_P = F^2(m_Q)
\left(1-\frac{4}{3}\,\frac{\alpha_s(m_Q)}{\pi}\right).
\end{equation}
At order $1/m_Q$, we find
\begin{eqnarray}
f_P \sqrt{m_P} & = &
F(m_Q)\left(1-\frac{2}{3}\,\frac{\alpha_s(m_Q)}{\pi}\right)\left[1+
\frac{1}{m_Q}\left\{-\frac{1}{2}\left(
1+\frac{8}{3} \,\frac{\alpha_s(m_Q)}{\pi}\right) \bar\Lambda +
\frac{1}{2}\,G_K(m_Q)\right.\right.\nonumber\\
& & \left.\left. + 3\left (1+\frac{13}{6}\,
\frac{\alpha_s(m_Q)}{\pi}\right) G_\Sigma(m_Q)\right\}\right].
\label{eq:fPexp}
\end{eqnarray}
From an analogous consideration of the vector current, it follows for
the vector decay constant $f_V$ \cite{N93}
\begin{eqnarray}
f_V\sqrt{m_V} & = & 
F(m_Q)\left(1-\frac{4}{3}\,\frac{\alpha_s(m_Q)}{\pi}\right)\left[1+
\frac{1}{m_Q}\left\{\frac{1}{6}\left(
1+\frac{4}{3} \,\frac{\alpha_s(m_Q)}{\pi}\right) \bar\Lambda +
\frac{1}{2}\,G_K(m_Q)\right.\right.\nonumber\\
& & \left.\left. - \left (1+\frac{13}{6}\,
\frac{\alpha_s(m_Q)}{\pi}\right) G_\Sigma(m_Q)\right\}\right].
\label{eq:fVexp}
\end{eqnarray}
These expressions are the generalization of (\ref{eq:deltaF}) to order
$\alpha_s$. Eq.\ (\ref{eq:fPexp}) coincides with the corresponding
expression derived in \cite{N93}.

The ratio of vector to pseudoscalar decay constant is then given by
\begin{equation}\label{eq:grr}
\frac{f_V}{f_P} = 1-\frac{2}{3}\,\frac{\alpha_s(m_Q)}{\pi} +
\frac{1}{m_Q}\left\{
\frac{2}{3}\,\bar\Lambda\,\left(1+\frac{5}{3}\,\frac{\alpha_s(m_Q)}{\pi}
\right) - 4 G_\Sigma (m_Q) \left( 1+
\frac{3}{2}\,\frac{\alpha_s(m_Q)}{\pi} \right)\right\}.
\end{equation}
From (\ref{eq:latest}) we now can construct the invariant quantities
\begin{eqnarray}
\hat{G}_K & = & G_K(\mu) - \bar{\Lambda}\left(
\frac{16}{3\beta_0}\,\ln \alpha_s(\mu) +
\frac{8}{3}\,T_{34}\,\frac{\alpha_s(\mu)}{\pi}\right),\nonumber\\
\hat{G}_\Sigma & = & [\alpha_s(\mu)]^{-3/\beta_0}\left(1- 
S_{\text{mag}}\,\frac{\alpha_s(\mu)}{4\pi}\right) \left\{
G_\Sigma(\mu) - \frac{8}{27}\,\bar{\Lambda}\left( 1+\frac{3}{2}\,T_{32}\,
\frac{\alpha_s(\mu)}{\pi}\right)\right\}.\label{eq:latest2}
\end{eqnarray}

\section{Sum rules for $G_K$ and $G_\Sigma$}

We now turn to the main part of this paper: the derivation of QCD sum
rules for $G_K$ and $G_\Sigma$. As mentioned in \mbox{Sec.\/ I}, a 
calculation of $G_K$ and $G_\Sigma$ from first principles is not
feasible to date, nor is it for hadronic matrix elements in general.
Instead one has to rely either on models invented {\em ad hoc} whose
relation to the underlying field theory remains unclear or one is
restricted to numerical approximations to QCD that allow the inclusion
of non--perturbative effects in a systematic way. One of these
appoximations is lattice QCD whose results can in principle be refined
to arbitrary accuracy, but in practice are limited by available
computer power and time. An other, less ambitious method
is provided by QCD sum rules (cf.\/ \cite{shifman} for 
reviews). Originally invented for the calculation of vacuum--to--meson
amplitudes in QCD \cite{SVZ},
it found application likewise to meson--to--meson transition
amplitudes \cite{ioffe} and could also be adapted to the study of HQET
\cite{BBBD,NA,BG92}. QCD sum rules rely on the field--theoretical
aspects and features of
QCD and were designed to make maximum use of known manifestations of
non--perturbative QCD. Their advantage over lattice calculations is
mainly their simplicity, whereas the main disadvantage may be seen in
the fact that a systematic refinement to a given accuracy is not
possible. The best accuracy one can hope to achieve is at the
$(10-20)\%$ level.

Without going too much into details, we want to present here only 
the gross 
features of the method. QCD sum rules are based on the operator
product expansion (OPE) of some correlation function, say
$\Pi(\omega)$, in terms of the so--called condensates, vacuum
expectation values of certain operators that vanish when taken over
the perturbative vacuum, but acquire non--zero values in the physical
QCD vacuum and thus characterize its non--perturbative nature. The
expansion is done in the not so deep Euclidean region of $\omega$
where non--perturbative effects are already noticeable, but do not
dominate yet:
\begin{equation}\label{eq:OPE}
\sum_n \Pi^{(n)}(\omega) \langle {\cal O}_n\rangle \equiv 
\sum_n \Pi^{(n)}(\omega) \langle 0|{\cal O}_n|0\rangle = \Pi(\omega).
\end{equation}
Examples for the vacuum condensates $\langle {\cal O}_n\rangle$
entering the OPE are $\langle {\cal O}_1\rangle=\openone$, the unity
operator, $\langle {\cal O}_3\rangle = \langle\bar{q} q\rangle$, the
(light) quark condensate, $\langle {\cal O}_4\rangle = \langle\alpha_s
G^2/\pi\rangle$, the gluon condensate, and $\langle {\cal O}_5\rangle =
\langle\bar{q} g_s\sigma_{\mu\nu} G^{\mu\nu}q\rangle$, the mixed
condensate. The coefficient functions $\Pi^{(n)}(\omega)$, also called
Wilson coefficients, can be
calculated within perturbation theory and are determined by the
short--distance behaviour of $\Pi(\omega)$.

On the other hand, $\Pi$ can be expressed via Cauchy's
theorem in terms of its singularities, which are situated on the
positive real axis in $\omega$ and are due to bound states causing 
poles and multi--particle states causing cuts. If we use as an input 
the information, that there {\em is} a pole due to the lowest--lying
one--particle state, and its position, we can fit its residue, the
square of the
vacuum--ground--state transition amplitude, from the knowledge of $\Pi$ in
the not so deep Euclidean region. The short--comings, however, of such an 
approach are obvious: first we have to take it for granted that the
OPE exists and makes sense. Besides, our knowledge of $\Pi$ in
the not so deep Euclidean region is not complete since one has to truncate
the OPE at some point, which leads to errors in the determination of the
residue. Moreover, there is not only the one--particle ground state,
but also radial excitations and multi--particle states contribute to 
the singularities of $\Pi$. Although it seems natural to assume that
contributions of multi--particle states be suppressed by creation 
amplitudes, one still has to deal with both the unwanted 
contributions from radial excitations and unknown terms in 
the OPE stemming from higher--dimensional condensates.

To this end, it proved extremely useful in many applications to 
subject $\Pi$ to a
Borel transformation, a tool, that likewise was introduced in
\cite{SVZ}. The Borel transform $\hat{B}F$ of a function $F$
depending on the Euclidean momentum $\omega_E=-\omega$ is defined as
\begin{equation}\renewcommand{\arraystretch}{0.7}
\hat{B}F = \lim_{\begin{array}{l} \scriptstyle \omega_E\to\infty,
N\to\infty\\\scriptstyle \omega_E/N=t\;\text{fixed}\\ \end{array}} 
\frac{1}{N!}\,(-\omega_E)^{N+1} \,\frac{d^{N+1}}{(d\omega_E)^{N+1}}\, F.
\end{equation}\renewcommand{\arraystretch}{1}
The dimensionful quantity replacing $\omega$ is $t$, the
Borel parameter. Applied to a dispersion relation over the physical 
spectral function $\rho^{\text{phys}}$, e.g., the Borel transformation
yields
\begin{equation}
\hat{B}\int\!\! ds\,\frac{\rho^{\text{phys}}(s)}{s-\omega} =
\frac{1}{t}\,\int\!\! ds\, \rho^{\text{phys}}(s)\,e^{-s/t}.
\end{equation}
Thus contributions of $\rho^{\text{phys}}(s)$ become exponentially
suppressed for large $s$, the suppression being controlled by the
value of $t$. Contributions of higher dimensional condensates to the
OPE (\ref{eq:OPE}) get suppression factors in powers of $1/t$ and
factorials. Roughly speaking, the Borel parameter serves as
weight between perturbative ($\propto\langle {\cal O}_1\rangle$) and
non--perturbative contributions.

Yet, one has to model $\rho^{\text{phys}}(s)$ for large values of $s$.
At that point the quark hadron duality hypothesis
enters. It states that integrals over $\rho^{\text{phys}}$ can be 
approximated to good accuracy by integrals over the perturbative
spectral function $\rho^{\text{pert}}$, multiplying $\langle 
{\cal O}_1\rangle$, to wit
\begin{equation}
\int\!\! ds\, \rho^{\text{phys}}(s)\, e^{-s/t} \approx \int\!\! ds\, 
\rho^{\text{pert}}(s)\,e^{-s/t}.
\end{equation}
Thus a suitable model for $\rho^{\text{phys}}$ for large values of $s$
is $\rho^{\text{phys}}(s) = \rho^{\text{pert}}(s)\Theta(s-\omega_0)$
where integration is understood afterwards. $\omega_0$ is the
so--called continuum threshold and is unfortunately not related to
fundamental quantities of QCD (or HQET), but should be less or equal
to the position of the lowest--lying excitation, i.e.\/
$\omega_0\approx 1\,\text{GeV}$. In the numerical analysis of sum
rules one thus has to look for a value of $\omega_0$ and a sufficiently
large range of $t$ where the sum rules becomes stable in these
parameters. The accuracy of about $(10-20)\%$ mentioned above can, 
however, be worsened by a possibly 
strong dependence of the result on the precise values of the Borel 
parameter and the continuum threshold. Stated
differently, too strong a dependence on that parameters is an
indication for the unreliabiliy of a sum rule. As we will see
in the next section, that is fortunately not the case with the
quantities we are interested in. 

Keeping all these caveats in mind, we proceed with the derivation of
QCD sum rules for $G_K$ and $G_\Sigma$. We write
\begin{equation}
T_K = \int\!\!ds\,\frac{\rho_K(s)}{s-\omega},\quad
T_\Sigma = \int\!\!ds\,\frac{\rho_\Sigma(s)}{s-\omega},
\end{equation}
do the OPE at the level of spectral densities and thus define
\begin{equation}
\rho_K = \rho_K^{\text{pert}} + \rho_K^{(3)}\langle\bar{q}q\rangle +
\rho_K^{(4)} \left\langle\frac{\alpha_s}{\pi}\,G^2\right\rangle +
\rho_K^{(5)} \langle\bar{q}g_s\sigma_{\mu\nu}G^{\mu\nu}q\rangle+\dots
\end{equation}
and correspondingly for $\rho_\Sigma$. We use Feynman gauge for
all diagrams involving only perturbative gluons; for the
calculation of non--perturbative diagrams we choose the
coordinate gauge \cite{fockschwinger}. The sign of the strong coupling
is defined by $D_\mu = \partial_\mu-ig_s A_\mu^A\lambda^A/2$. We apply
dimensional regularization in $D=4+2\epsilon$ dimensions and use the
$\overline{\text{MS}}$ renormalization scheme.
For the perturbative contribution, we take into account the diagrams
shown in Fig.\ \ref{fig:diagspert} and find (at the renormalization
scale $\mu$):
\begin{eqnarray}
\rho_K^{\text{pert}} & = & \frac{s^3}{\pi^2}\left\{-3+
\frac{\alpha_s(\mu)}{\pi}\left(-\frac{137}{6}- \frac{4}{3}\pi^2 +
 10 \ln\frac{2s}{\mu}\right)\right\}\Theta(s),\nonumber\\
\rho_\Sigma^{\text{pert}} & = & \frac{s^3}{\pi^2}\,
\frac{\alpha_s(\mu)}{\pi}\left(\frac{7}{9}-\frac{4}{3}\,\ln
\frac{2s}{\mu} \right)\Theta(s).
\end{eqnarray}
The loop--integrals involved in the calculation can be found in
\cite{BB93}, e.g.
The diagrams for the contribution of the quark condensate are
shown in Fig.\ \ref{fig:diagsquark} and yield
\begin{equation}
\rho_K^{(3)} = 0,\quad
\rho_\Sigma^{(3)} = \frac{2}{3}\,\frac{\alpha_s(\mu)}{\pi}\,\Theta(s).
\end{equation}
In addition we have taken into account the contributions of the gluon
and the mixed condensate shown in \mbox{Fig.\/ \ref{fig:diags45dim}}
that read
\begin{eqnarray}
\rho_K^{(4)} &= & -\frac{1}{48}\,\delta(s),\quad
\rho_K^{(5)} = \frac{3}{32}\,\delta'(s),\nonumber\\
\rho_\Sigma^{(4)} & = & \frac{1}{48}\,\delta(s),\quad
\rho_\Sigma^{(5)} = \frac{1}{48}\,\delta'(s).
\end{eqnarray}
From the result for $\Pi(\omega)$ given in \cite{BBBD,NA,BG92} we obtain
\begin{equation}
\rho_{\omega\Pi} = \frac{3}{2\pi^2}\,s^2\left\{ 1+
\frac{\alpha_s(\mu)}{\pi} \left(\frac{17}{3}+\frac{4}{9}\pi^2 -2\ln
\frac{2s}{\mu} \right)\right\} \Theta(s) - \frac{1}{16}\, \delta(s)\,
\langle\bar{q}g_s\sigma_{\mu\nu}G^{\mu\nu}q\rangle.
\end{equation}
We can now check all the formul\ae\/ obtained so
far by means of Eq.\ (\ref{eq:expansion}). The expression for $\Pi_5$ 
including non--perturbative corrections
can be found in \cite{fBQCD}, e.g. Here we concentrate on the
perturbative part calculated in \cite{broadhurst}. For its expansion to
$O(1/m_Q)$, we recall that (\ref{eq:expansion}) is valid up
to constants, so we make frequently use of partial integration to
obtain the spectral density. Expanding the four--momentum as
$q^2=m_Q^2 + 2m_Q \omega+\omega^2$, 
we find\footnote{For the expansion of the full $\Pi_5$, note that the 
radiative correction to the Wilson
coefficient multiplying the quark condensate was first calculated 
in \cite{BBBD,NA} and reads before application of the Borel
transformation:
$$
\Pi_5^{(3)}\langle\bar{q}q\rangle = (m_{\overline{\text{MS}}} \langle
\bar{q}q\rangle)\,
\frac{m_Q^2}{q^2-m_Q^2}\,\left\{ 1+ 2\,\frac{\alpha_s}{\pi} \left( 1+
\frac{q^2-m_Q^2}{q^2}\, \ln \frac{m_Q^2}{m_Q^2-q^2}\right)\right\}.
$$}
\begin{eqnarray}\label{eq:latest3}
\rho_{\Pi_5}^{\text{pert}} & = & m_Q^2 \left[ \frac{3}{2\pi^2}\,s^2\left\{ 1+
\frac{\alpha_s}{\pi}\left( \frac{13}{3} - 2\ln\frac{2s}{m_Q} +
\frac{4}{9}\,\pi^2 \right)\right\}\right.\nonumber\\
& & \phantom{m_Q^2\:}\left. + \frac{1}{m_Q}\,\frac{s^3}{\pi^2} 
\left\{ -\frac{3}{2} + \frac{\alpha_s}{\pi} \left( -14 +
3\ln\frac{2\omega}{m_Q} -\frac{2}{3}\,\pi^2\right)\right\}\right]
\Theta(s).
\end{eqnarray} 
Inserting $C(\mu)$, $B_1(\mu)$ and $B_2(\mu)$ from (\ref{eq:PScoeff})
and $\rho_{\omega\Pi}^{\text{pert}}$, $\rho_K^{\text{pert}}$ and 
$\rho_\Sigma^{\text{pert}}$ from above into the right hand side of 
(\ref{eq:expansion}), we indeed recover (\ref{eq:latest3}).

There remains only one slight complication to cope with, that is the
structure of $T_K$ and $T_\Sigma$ expressed in terms of hadronic
matrix elements. As mentioned at the end of the last section, both
quantities develop apart from the single poles at $\omega=\bar\Lambda$
we are interested in also double poles that stem from the contribution
of three--point functions\footnote{Such a mixing of two-- and
three--point functions was first considered in \cite{BK87}.} and can
be obtained by inserting hadronic
states between each of the operators in $\langle J_{\Gamma_1}{\cal O}
J_{\Gamma_2}\rangle$. With $K = \langle M|{\cal K}|M\rangle/(2m_M)$ the
invariant quantity $\hat{T}_K$ defined in (\ref{eq:latest}) can be 
expressed as
\begin{equation}
\hat{T}_K = \frac{\hat{F}^2 \hat{G}_K}{2(\bar\Lambda-\omega)} +
\frac{\hat{F}^2 \hat{K}}{4(\bar\Lambda-\omega)^2}+\,\text{higher states.}
\end{equation}
We can however isolate the single pole contribution by first applying 
a Borel transformation,
\begin{eqnarray}
\hat{B}\hat{T}_K & = & \frac{1}{4t^2}\, \hat{F}^2\hat{K}e^{-\bar\Lambda/ t} +
\frac{1}{2t}\, \hat{F}^2\hat{G}_K e^{-\bar\Lambda/t} + \dots\nonumber\\
& = & \frac{1}{2t}\,\hat{K}\hat{B}\hat{\Pi} +
\hat{G}_K\hat{B}\hat{\Pi}+\dots,
\end{eqnarray}
where in the step from the first to the second equation we have made
use of the Borel transform of Eq.\/ (\ref{eq:xyz}). Taking then the
derivative with respect to $t$ yields
\begin{equation}\label{eq:x1}
\hat{G}_K =
\frac{d}{dt}\,\frac{t\hat{B}\hat{T}_K}{\hat{B}\hat{\Pi}}.
\end{equation}
The corresponding equation for $\hat{G}_\Sigma$ is
\begin{equation}\label{eq:x2}
\hat{G}_\Sigma = \frac{1}{2}\,\frac{d}{dt}\, 
\frac{t\hat{B}\hat{T}_\Sigma}{\hat{B}\hat{\Pi}}.
\end{equation}
The residues of the double pole contributions were calculated in 
\cite{BB93}. Eqs.\/ (\ref{eq:x1}) and (\ref{eq:x2}) are
the final sum rules we are going to evaluate in the next section.

Let us close with a few remarks about the calculation 
performed in \cite{N92}. In our approach, the contributions of $O(1)$ are
completely separated from the ones of $O(1/m_Q)$, because the matrix
elements considered are different. So from our point of view it is not
allowed to introduce a shift of the Borel parameter of order $1/m_Q$
in the sum rule for $f_M$ as it was done in \cite{N92} since this would mix 
different orders in the inverse heavy quark mass. Since we have
derived sum rules for $G_K$ and $G_\Sigma$ separately, there is no
need to shift the value of the continuum threshold, either.
Furthermore, the double pole contributions escaped the attention of
Ref.\ \cite{N92} and thus the results obtained there are not values
for $G_K$ and $G_\Sigma$, but mixtures of $G_K$ and $K$ and of 
$G_\Sigma$ and $\langle M|{\cal S}|M\rangle$.
 
\section{Results and Discussion}

In the numerical evaluation of the sum rules (\ref{eq:x1}) and
(\ref{eq:x2}) we use the following values of the condensates at the
renormalization point $\mu=1\,\text{GeV}$ \cite{shifman}:
\begin{eqnarray}
& & \langle\bar{q}q\rangle = (-0.24\,\text{GeV})^3,\quad \left\langle
\frac{\alpha_s}{\pi}\,G^2\right\rangle =
0.012\,\text{GeV}^4,\nonumber\\
& & \langle\bar{q}g_s\sigma_{\mu\nu}G^{\mu\nu}q\rangle =
0.8\,\text{GeV}^2\, \langle\bar{q}q\rangle.
\end{eqnarray}
For scales different from 1 GeV, we scale the condensates accordingly.
For $\alpha_s$, we use
\begin{equation}
\alpha_s(\mu) =\frac{4\pi}{2\beta_0\ln(\mu/\bar\Lambda^{(n_f)}_{ 
\overline{\text{MS}}})}\left[1-\frac{\beta_1}{\beta_0^2}\, \frac{\ln
(2\ln (\mu/\bar\Lambda^{(n_f)}_{\overline{\text{MS}}}))}{2
\ln(\mu/\bar\Lambda^{(n_f)}_{\overline{\text{MS}}})}\right]
\end{equation}
with $\bar\Lambda^{(4)}_{\overline{\text{MS}}} = 260\,\text{MeV}$,
corresponding to $\alpha_s(5\,\text{GeV}) = 0.20$. 

\begin{figure}[h]
\vspace{-0.6in}
\centerline{
\epsfbox{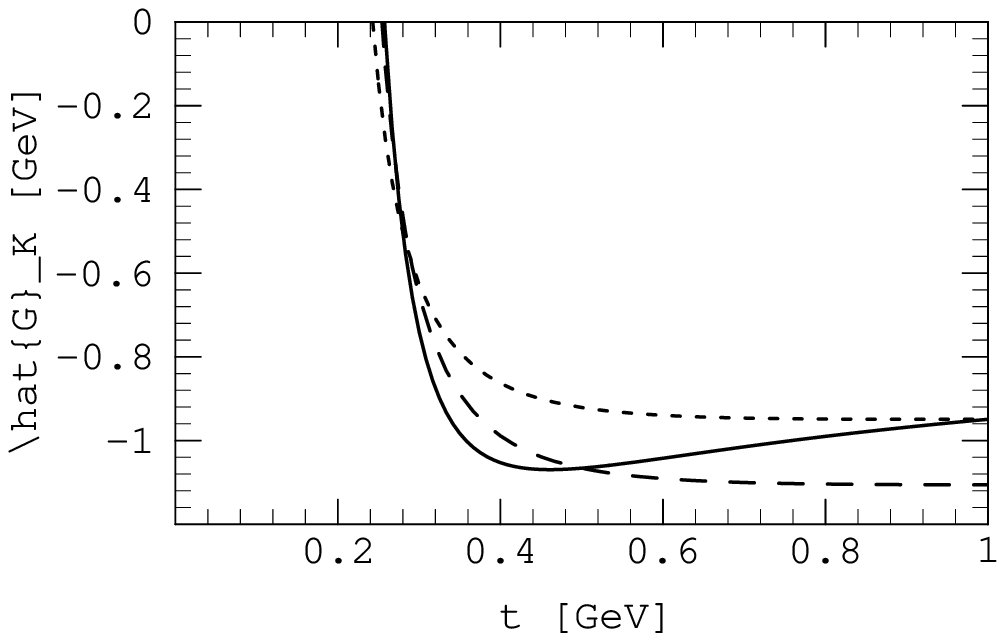}}
\vspace{-0.7in}
\caption[]{$\hat{G}_K$ as function of $t$ for $\omega_0=1\,\text{GeV}$
and different scales $\mu$. Solid line: $\mu = 2t$, long--dashed 
line: $\mu = 1\,\text{GeV}$, short--dashed line: $\mu = 
2\,\text{GeV}$.}\label{fig:6}
\vspace{-0.6in}
\centerline{
\epsfbox{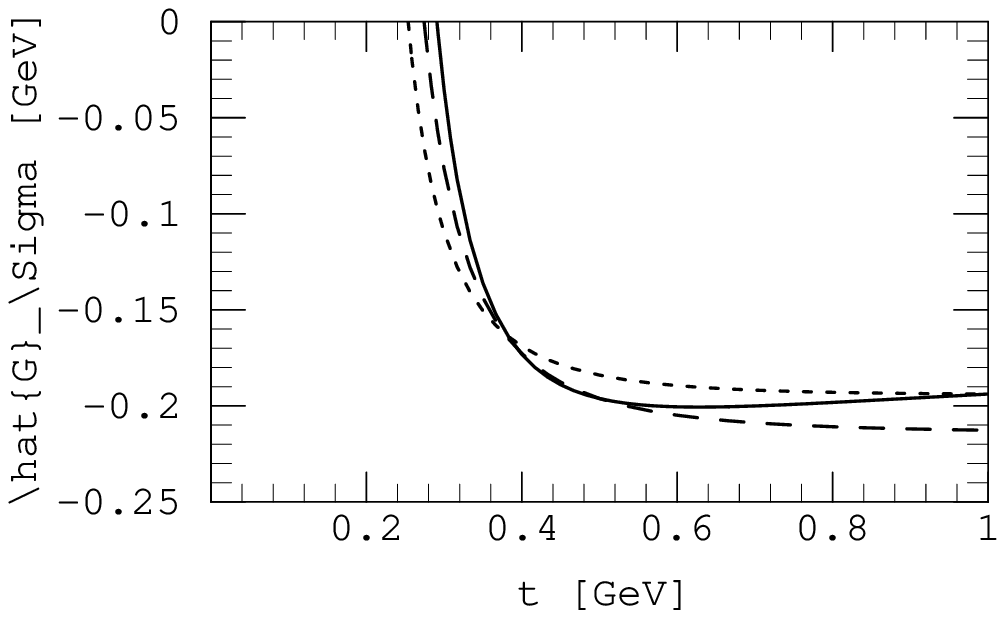}}
\vspace{-0.7in}
\caption[]{$\hat{G}_\Sigma$ as function of $t$ for $\omega_0=1\,\text{GeV}$
and different scales $\mu$. Solid line: $\mu = 2t$, long--dashed 
line: $\mu = 1\,\text{GeV}$, short--dashed line: $\mu = 
2\,\text{GeV}$}\label{fig:7}
\vspace{-0.6in}
\centerline{
\epsfbox{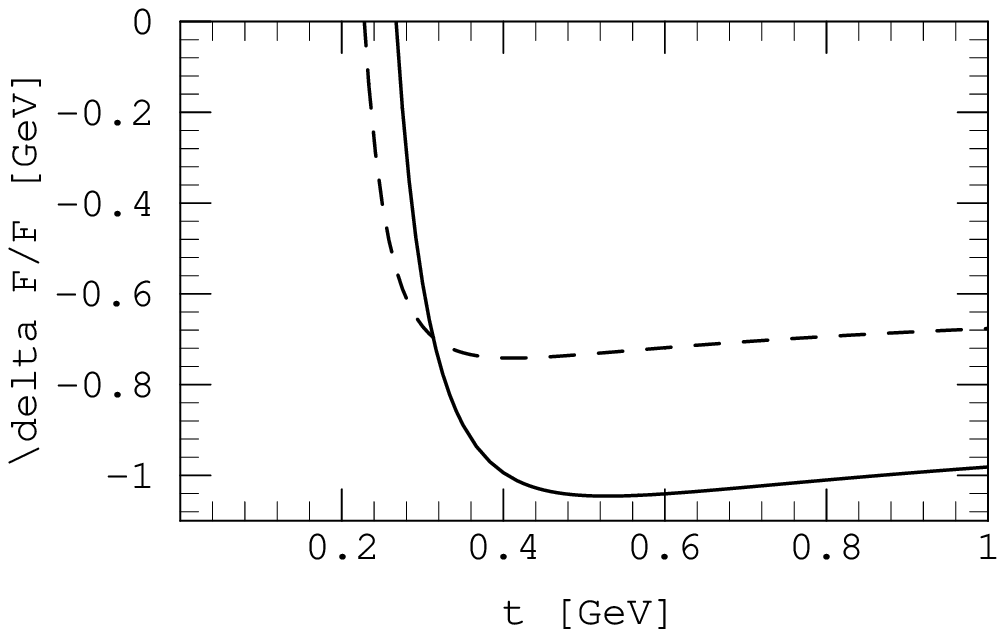}}
\vspace{-0.7in}
\caption[]{The $1/m_Q$ corrections $\delta F/F$ to $F$ according to
Eqs.\ (\protect{\ref{eq:fPexp}}) and (\protect{\ref{eq:fVexp}}) as 
function of $t$ for
$\omega_0=1\,\text{GeV}$. Solid line: pseudoscalar decay constant,
dashed line: vector decay constant. The normalization scale is $\mu =
2t$.}\label{fig:8}
\end{figure}

In Fig.\ \ref{fig:6} we show $\hat{G}_K$ as function of $t$ for
$\omega_0=1\,\text{GeV}$. Although $\hat{T}_K$ is renormalization
group invariant to next--to--leading order, we still have to choose
one particular scale in building this quantity (cf.\ (\ref{eq:latest})).
As argued in \cite{BBBD}, the most natural choice is $\mu=2t$. In
Fig.\ \ref{fig:6} we also show curves for $\mu=1\,\text{GeV}$ and
$\mu=2\,\text{GeV}$. The scale--dependence still visible indicates the
size of higher order effects in $\alpha_s$. $\hat{T}_K$ also depends
on the unknown anomalous dimension matrix element $T_{34}$ whose value
is put zero in the figure. Varying the continuum threshold within
$\omega_0 = (1.0-1.2)\,\text{GeV}$ and $t$ within 
$0.35\,\text{GeV}<t<0.6\,\text{GeV}$, as suggested by the analysis for
$F$ done in \cite{BBBD}, and $T_{34}$ in the range from $-1$ to $1$, we
obtain
\begin{equation}
\hat{G}_K = -(1.1\pm 0.2\pm 0.2)\,\text{GeV}.
\end{equation}
The first error indicates the sum rule uncertainty in the ``sum rule
window'' $0.35\,\text{GeV}<t<0.6\,\text{GeV}$, the second one the
uncertainty stemming from the variation of $T_{34}$ 
which is rather large. From (\ref{eq:latest2}) we then obtain
\begin{equation}\label{eq:GKmB}
G_K(m_Q) = -(1.6\pm 0.2\pm 0.3\pm 0.1)\,\text{GeV},
\end{equation}
where the first two errors have the same meaning as before, the third
one stems from varying $\bar\Lambda$ in the range
$(0.5-0.6)\,\text{GeV}$.

In Fig.\ \ref{fig:7}, $\hat{G}_\Sigma$ is displayed as function of $t$
for $\omega_0=1\,\text{GeV}$. As for $\hat{G}_K$, we have varied
$\mu$ from 1 to 2 GeV. Although here two unknown matrix
elements enter, $S_{\text{mag}}$ and $T_{32}$ whose values we have put
to zero
in the figure, their influence on the result is much smaller than that
of $T_{34}$ on $G_K$ and we
find
\begin{equation}
\hat{G}_\Sigma = -(0.19\pm0.06\pm0.02)\,\text{GeV}.
\end{equation}
which corresponds to
\begin{equation}\label{eq:GSmB}
G_\Sigma(m_Q) = (0.042\pm0.034\pm0.023\pm0.030)\,\text{GeV}.
\end{equation}
The large errors in this result come from a cancellation of the
central values of the two
terms in (\ref{eq:latest2}). 

In the last figure, Fig.\ \ref{fig:8}, we show the resulting $1/m_Q$
corrections for $f_P\sqrt{m_P}$ and $f_V\sqrt{m_V}$, denoted as
$\delta F/F$, that are obtained from Eqs.\ (\ref{eq:fPexp}) and 
(\ref{eq:fVexp}). They yield\footnote{Note that
the errors of $G_\Sigma(m_Q)$ and $G_K(m_Q)$ are correlated.}
\begin{eqnarray}
f_P\sqrt{m_P} & = &
F(m_Q)\left(1-\frac{2}{3}\,\frac{\alpha_s(m_Q)}{\pi}\right) \left( 1 -
\frac{1}{m_Q}\, (1.07\pm 0.19\pm 0.07 \pm 0.02)\,\text{GeV}\right),
\nonumber\\
f_V\sqrt{m_V} & = &
F(m_Q)\left(1-\frac{4}{3}\,\frac{\alpha_s(m_Q)}{\pi}\right) \left( 1 -
\frac{1}{m_Q}\, (0.78\pm 0.08\pm 0.09 \pm 0.04)\,\text{GeV}\right)
\label{eq:zzz}
\end{eqnarray} 
with the same sequence of errors as before. Note that the error
stemming from the unknown matrix elements is reduced as compared with
(\ref{eq:GKmB}) and (\ref{eq:GSmB}) and that the sum rule uncertainty
for $f_V$ is smaller than that for $f_P$ due to the better stability
visible in Fig.\ \ref{fig:8}.

Finally, we obtain from (\ref{eq:grr}) (at the scale $m_b$)
\begin{equation}
\frac{f_V}{f_P} = 1.00\pm 0.03\pm 0.02\pm 0.02 + O(1/m_b^2),
\end{equation}
that is, a nearly cancellation of the $1/m_Q$ term, indicating that
 $1/m_Q^2$ correction terms become important.

The result (\ref{eq:zzz}) agrees well with our
old one obtained from the QCD sum rule for $f_P$ \cite{BBBD}:
\begin{equation}
f_P\sqrt{m_P} = F(m_Q) \left(1-\frac{2}{3}\,\frac{\alpha_s(m_Q)}{\pi}
\right) \left(1-\frac{(0.8-1.1)\,\text{GeV}}{m_Q}\right).
\end{equation}
The remaining slight discrepancy is within the uncertainty of the
method and/or could be due to $1/m_Q^2$ correction terms. Our result
also agrees well with values obtained from lattice calculations
\cite{fBlatt}. As far as \cite{N92} is concerned, our value for
$G_K(m_Q)$ is by nearly a factor three smaller, for $G_\Sigma(m_Q)$ we
obtain a different sign. The difference in sign affects mainly the
ratio $f_V/f_P$, where we find some disagreement with the lattice
calculation \cite{lattat} where $f_V/f_P$ is found to be $1.11\pm 
0.05$ at the scale of the B meson. This value, however, relies on a
linear fit through
four points in the mass range $(1.3-2.6)\,\text{GeV}$ that has to be
extrapolated to a mass of 5.28 GeV. Fitting a parabel reduces the
ratio already to 1.06 instead of 1.11. The remaining discrepancy could
be due to $1/m_Q^2$ terms that mimic a $1/m_Q$ behaviour in the
restricted mass range considered in \cite{lattat}.

Concluding, we have found the sum rules for $1/m_Q$ corrections in
HQET in complete agreement with conventional
sum rules. There is however, a discrepancy in $f_V/f_P$ compared
to lattice calculations which might be due to
$1/m_Q^2$ correction terms.

\acknowledgments
It is a pleasure to thank V.M.\ Braun for useful discussions and
collaboration in the early stages of this work. This work was supported
in part by the German Bundesministerium f\"ur Forschung und Technologie
under contract 06 TM 732 and by the CEC science project
SC1--CT91--0729.

\begin{figure}[h]
\vspace{0.3in}
\centerline{
\epsfbox{diagmom.ps}}
\vspace{0.3in}
\caption[]{Momentum configuration for the matching calculation of the
pseudoscalar current. $p$ is the momentum of the heavy quark $Q$, $p'$
that of the massless quark $q$. Both quarks are 
on--shell.}\label{fig:momconf}
\vspace{0.3in}
\centerline{
\epsfbox{diagspert.ps}}
\vspace{0.3in}
\caption[]{Perturbative contribution to $T_K$. Double lines denote
heavy quarks, single lines light quarks and curly lines gluons. The
crosses with circles denote the heavy--light vertex $\Gamma$, the
black square the insertion of the operator ${\cal K}$. All diagrams
not drawn vanish.}\label{fig:diagspert}
\vspace{0.3in}
\centerline{
\epsfbox{diagsquark.ps}}
\vspace{0.3in}
\caption[]{Diagrams contributing to the Wilson coefficient of the
quark condensate in the OPE of $T_K$. Lines with crosses denote vacuum
expectation values.}\label{fig:diagsquark}
\vspace{0.3in}
\centerline{
\epsfbox{diags45dim.ps}}
\vspace{0.3in}
\caption[]{Diagrams contributing to the Wilson coefficient of the
gluon and the mixed condensate in the OPE of $T_K$. We use the 
coordinate gauge with the coordinates specified in the second
diagram. All diagrams not drawn vanish.}\label{fig:diags45dim}
\vspace{0.3in}
\centerline{
\epsfbox{diagsSigma.ps}}
\vspace{0.3in}
\caption[]{Diagrams contributing to $T_\Sigma$. The black square now
denotes an insertion of the operator ${\cal
S}$.}\label{fig:diagsSigma}
\end{figure}
\clearpage

\makebox[1cm]{}

\vspace{-1.1in}

\end{document}